\documentclass[12pt]{article}

\usepackage{amsmath,amssymb,amsthm,mathrsfs,bbm}
\usepackage{latexsym,amscd,amsbsy,dsfont,amsfonts}
\usepackage{graphicx}
\usepackage[pdftex,colorlinks=true]{hyperref}

\newcommand{\hs}{\hspace{0.15mm}}

\newcommand{\bi}{\int_{\partial M}}

\usepackage{bm}
\usepackage{latexsym}
\usepackage[latin1]{inputenc}
\usepackage{amsfonts}
\usepackage{amssymb}
\usepackage{amsmath}
\usepackage{subfigure}

\newcommand{\be}{\begin{equation}}
\newcommand{\ee}{\end{equation}}
\newcommand{\bea}{\begin{eqnarray}}
\newcommand{\eea}{\end{eqnarray}}

\begin{document}

\makeatletter
\renewcommand{\theequation}{\thesection.\arabic{equation}}
\@addtoreset{equation}{section}
\makeatother

\baselineskip 18pt

\begin{titlepage}

\vfill

\begin{flushright}
\end{flushright}

\vfill

\begin{center}
   \baselineskip=16pt
   {\Large\bf  A simple holographic model of momentum relaxation}
  \vskip 1.5cm
      Tom\'as Andrade$^{1}$ and Benjamin Withers$^{2}$\\
   \vskip .6cm
        \begin{small}
      {}$^{1}$\textit{Centre for Particle Theory, Department of Mathematical Sciences\\
  Durham University, South Road, Durham DH1 3LE, U.K.}
        \vskip2em
        {}$^{2}$\textit{Mathematical Sciences and STAG Research Centre\\
         University of Southampton, Highfield, Southampton SO17 1BJ, U.K.}
        \end{small}\\*[.6cm]
   \end{center}

\vfill

\begin{center}
\textbf{Abstract}
\end{center}

We consider a holographic model consisting of Einstein-Maxwell theory in $d+1$ bulk spacetime dimensions with $d-1$ massless scalar fields. Momentum relaxation is realised simply through spatially dependent sources for operators dual to the neutral scalars, which can be engineered so that the bulk stress tensor and resulting black brane geometry are homogeneous and isotropic. We analytically calculate the DC conductivity, which is finite.  In the $d=3$ case, both the black hole geometry and shear-mode current-current correlators are those of a sector of massive gravity. 

\vfill

\end{titlepage}
\setcounter{equation}{0}

\section{Introduction}
Consider a $d+1$-dimensional bulk gravitational model containing a scalar field $\psi$ dual to an operator with a spatially dependent source, $\psi^{(0)}(x^i)$. In this case, the holographic stress tensor is not divergence free, but instead obeys a conservation equation with contributions from the scalar vev, $\langle O \rangle$,
\begin{equation}\label{ward intro}
	\nabla_i \langle T^{ij} \rangle = \nabla^j \psi^{(0)} \langle O \rangle + F^{(0)ji} \langle J_i \rangle,
\end{equation}
where $i,j$ labels the boundary spacetime directions, see e.g. \cite{Bianchi:2001kw}. The last term is present if we also turn on a boundary field strength $F^{(0)ji} = d A^{(0)}$, with current $\langle J_i \rangle$. The Ward identity \eqref{ward intro} suggests a route to holographic momentum relaxation and finite DC optical conductivity by turning on a spatially dependent source term for the scalar. 

Spatially dependent sources have been utilised to construct holographic lattices \cite{Horowitz:2012ky,Horowitz:2012gs,Horowitz:2013jaa,Ling:2013nxa,Chesler:2013qla,Ishibashi:2013nsa} which exhibit finite DC conductivity, in agreement with ionic lattice computations of \cite{Hartnoll:2012rj}. However, one is not restricted to explicit lattice configurations.  Momentum relaxation has also been investigated in holography by making use of neutral matter which acts as a large reservoir into which momentum can be dissipated by charge carriers which are treated in the probe approximation \cite{Karch:2007pd,Faulkner:2010da,Hartnoll:2009ns}.


In a separate line of attack, approaches to momentum relaxation have been made by considering a theory which explicitly breaks diffeomorphism invariance in the bulk. Specifically the use of non-linear massive gravity theory of \cite{deRham:2010kj} was initiated in the holographic context by \cite{Vegh:2013sk}, and explored further in \cite{Davison:2013jba,Blake:2013bqa}. In these cases it was found that the DC conductivity is indeed finite, with Drude physics resulting \cite{Davison:2013jba}. However, the field theory interpretation of massive gravity is at present unclear. It was suggested in \cite{Blake:2013owa} that a particular theory of massive gravity is related to lattice physics.

The point of this paper is to investigate the role of spatially dependent field theory sources on momentum relaxation in the simplest possible way. Namely, we avoid the scenario where the stress tensor depends on the boundary coordinates. To achieve this we exploit a scalar field shift symmetry. Specifically we make $\psi$ massless, so that it only enters the bulk stress tensor through $\partial_\mu \psi$, and we only turn on sources which are linear in the boundary coordinates, 
\be
\psi^{(0)} \propto \alpha_i x^i. 
\ee
Consequentially we can find homogeneous bulk solutions.  In general though, such a configuration will not be isotropic. To simplify things further we include a total of $d-1$ scalar fields, $\psi_I$, where $d-1$ corresponds to the number of spatial dimensions of the boundary theory. With this number of scalar fields, we can arrange their sources, $\alpha_{Ii} x^i$ such that the bulk metric is also isotropic. 

With these field theory sources, we analytically obtain finite temperature charged black brane solutions. We calculate the DC conductivity of these black holes in general dimensions, finding,
\be\label{sigma DC intro}
	\sigma_{DC} = r_0^{d-3}\left( 1 + (d-2)^2 \frac{\mu^2}{\alpha^2}\right).
\ee
where $\alpha$ is a parameter introduced later and is given by the gradient of the source terms, $\mu$ is the chemical potential and $r_0$ the black brane horizon position.

A key result of this work is that in the case of four dimensions ($d=3$) both the background black brane solutions and the shear-mode current-current correlators are the same as those of a sector of holographic non-linear massive gravity. Hence we have an equivalent description of much of the same physics investigated in \cite{Vegh:2013sk,Davison:2013jba,Blake:2013bqa}, but in a holographic context where the field theory origin is better understood.

In this paper we employ spatially dependent configurations in such a way as to retain bulk homogeneity. This is similar in spirit to constructions in 5 bulk dimensions where Bianchi VII$_0$ symmetry can be exploited to construct helical black holes through the solution of ODEs rather than PDEs \cite{Donos:2012js}. Using a spatially dependent phase for a complex scalar field has been proposed in this context, giving again ODEs in the bulk. This possibility was mentioned in \cite{Blake:2013owa,Donos:2013eha} and explored in some detail in \cite{Donos:2013eha}, which appeared as this paper was being finalised.

The paper is organised as follows.  We begin with the $d+1$ dimensional black brane solutions including the non-trivial scalar sources in section \ref{blackbranes}, and calculate their associated DC conductivity in section \ref{sec conductivity}. In section \ref{sec D=4} we specialise to the case of four bulk spacetime dimensions, $d=3$, where we can perform in detail the holographic renormalisation procedure enabling us in section \ref{sec: comp MG} to make the connection with massive gravity. We conclude in section \ref{discussion}.

\section{Black branes with scalar sources
\label{blackbranes}}

The model we use throughout this paper is,
\begin{equation}
	S_0 = \int_M \sqrt{-g} \left[ R - 2 \Lambda - \frac{1}{2} \sum_{I}^{d-1} (\partial \psi_I)^2  - \frac{1}{4} F^2 \right ] d^{d+1} x 
	- 2 \int_{\partial M} \sqrt{-\gamma} K d^dx. \label{the model}
\end{equation}
where $\Lambda = - d(d-1)/(2 \ell^2)$, and we have included the field strength $F=dA$ for a $U(1)$ gauge field $A$, and the $d-1$ massless scalar fields, $\psi_I$. We have chosen units such that the gravitational coupling satisfies $16\pi G=1$. In \eqref{the model}, $\gamma$ is the induced metric on the boundary and $K$ is the trace of the extrinsic curvature, given by $K_{\mu\nu} = -\gamma_{\mu}^\rho \gamma_{\nu}^\sigma\nabla_{(\rho}n_{\sigma)}$ where $n$ is the outward pointing unit normal to the boundary.

The model \eqref{the model} admits AdS$_{d+1}$ solutions with AdS radius $\ell$. Henceforth we will set $\ell = 1$. It also
admits homogeneous and isotropic charged black brane solutions with non-trivial scalar field sources, whose properties will be the focus of this paper. They can be constructed analytically. Working in a coordinate system in which $r \to \infty$ labels the UV boundary, these solutions have the form     
\begin{equation}\label{gen D ansatz}
	ds^2 = - f(r) dt^2 + \frac{dr^2}{f(r)} + r^2 \delta_{ab} dx^a dx^b,  \qquad A = A_t(r) dt,  \qquad \psi_I = \alpha_{Ia} x^a, 
\end{equation}
where $a$ labels the $d-1$ spatial $x^a$ directions, $I$ is an internal index that labels the $d-1$ scalar fields and $\alpha_{Ia}$ are real arbitrary constants. 

It is a simple exercise to verify that the anstaz \eqref{gen D ansatz} yields a solution of the equations of motion if 
 \begin{align}
\label{f gen d}
	f &= r^2  - \frac{\alpha^2}{2(d-2)} - \frac{m_0}{r^{d-2}} + \frac{(d-2) \mu^2 }{2 (d-1)} \frac{ r_0^{2(d-2)} }{ r^{2(d-2)} },  \\
\label{At soln gen d}
	A_t &= \mu \left(1 - \frac{r_0^{d-2}}{r^{d-2}}\right),
\end{align}
where 
\begin{equation}
	\alpha^2 \equiv \frac{1}{d-1}\sum_{a= 1}^{d-1} \vec\alpha_a \cdot \vec \alpha_a,  
\end{equation}
provided
\begin{equation}
	\vec\alpha_a \cdot \vec \alpha_b = \alpha^2 \delta_{ab}\quad\quad \forall a, b.
\end{equation}
Here we have introduced the vector notation $(\vec \alpha_a)_I = \alpha_{I a} $ and $\vec\alpha_a \cdot \vec \alpha_b = \sum_I \alpha_{Ia} \alpha_{Ib} $. These and related solutions have previously been explored in \cite{Bardoux:2012aw}, additionally see \cite{Iizuka:2012wt} for anisotropic examples with a single scalar.

The parameter $\mu$ is interpreted as the chemical potential in the dual field theory, while $m_0$ is proportional to the energy density of the brane and may be obtained by solving  $f(r_0)=0$ so that $r_0$ gives the horizon location,
\be
m_0 = r_0^d \left(1+ \frac{d-2}{2(d-1)}\frac{\mu^2}{r_0^2}- \frac{1}{2(d-2)}\frac{\alpha^2}{r_0^2} \right).
\ee
Note that in writing \eqref{At soln gen d} we have chosen a gauge in which $A$ vanishes at the horizon. The temperature of the black hole is given by
\begin{equation}\label{T gen d}
	T = \frac{f'(r_0)}{4 \pi} = \frac{1}{4 \pi} \left( d r_0 - \frac{\alpha^2}{2r_0}  -  \frac{(d-2)^2\mu^2}{2 (d-1) r_0}   \right). 
\end{equation}
We are working in units for which $1 6 \pi G = 1$, so the entropy density $s$ can be written as
\begin{equation}\label{s gen d}
	s = 4 \pi r_0^{d-1}.
\end{equation}

At $T=0$ the black hole becomes a finite entropy domain wall which interpolates between unit-radius AdS$_{d+1}$ in the UV and a near horizon AdS$_2 \times \mathbb{R}^{d-1}$, where the AdS$_2$ radius, $\ell_{\text{AdS}_2}$, is given by
\be
\ell_{\text{AdS}_2}^2 =\frac{1}{d(d-1)} \frac{(d-1)\alpha^2 + (d-2)^2\mu^2}{\alpha^2 + (d-2)^2\mu^2}.
\ee
Note that this near horizon geometry can be obtained even when $\mu=0$, supported by $\alpha$.

Summing up, for a given mass and chemical potential, we have a $(d-1)$-parameter solution, which we can choose to label 
by one of the $\vec \alpha_a$. Note however that there is a redundancy associated to the rotational symmetry on the $x^a$ space, 
which we can used to eliminate all but one of these free parameters. Therefore, the solution is fully specified by $T$, $\mu$ and $\alpha$.  

\section{Conductivity\label{sec conductivity}}


The main feature of the model \eqref{the model} is that it is suitable to describe momentum dissipation in the dual field theory. As mentioned above, this occurs due to spatially dependent field theory sources. To see this, we recall that, in the presence of both scalar and gauge field sources, the Ward identity for the (non)-conservation of the boundary  stress tensor is 
\begin{equation}\label{ward cons general}
	\nabla_i \langle T^{ij} \rangle = \nabla^j \psi_I^{(0)} \langle O_I \rangle + F^{(0)ji} \langle J_i \rangle ,
\end{equation}
\noindent where $i, j$ are indices in the field theory $x^i = (t, x^a$), $\psi^{(0)}$ is the source for the operator $\langle O \rangle$, $F^{(0)}_{ij}$ is the field strength associated to a background gauge field $A^{(0)}_i$ which couples to a conserved $U(1)$ current. 
It is clear from \eqref{ward cons general} that turning on sources in the presence of non-zero vevs yields the non-conservation of $\langle P^a \rangle \equiv \langle T^{ta} \rangle$. 

We will postpone an explicit holographic derivation of \eqref{ward cons general} in \eqref{the model} specialized to four bulk dimensions until section \ref{sec D=4}. For the moment we will explore its consequences in the context of linear fluctuations. In particular, we shall see that the DC conductivity in our model is finite, as a consequence of the lack of conservation just described.  Moreover, we shall be able to compute the value of the DC conductivity for arbitrary values of the parameters. 

In order to compute the conductivity, we consider linearized fluctuations of the form 
\begin{equation}
	 \delta A_x = e^{- i \omega t} a_{x}(r),  \qquad \delta g_{tx} = e^{- i \omega t} r^2 H_{tx}(r)
\end{equation}
\noindent and all the other metric and gauge perturbations vanishing. Here $x$ is one of the spatial boundary indices. In order to simplify the scalar equations, we use the residual symmetry of the background to set $\alpha_{Ia} = \alpha \delta_{Ia}$, so that, without loss of generality, we take every background scalar field to have only one non-zero contribution that is linear in the boundary coordinates.

In this set up, it is consistent to set all scalar fluctuations to zero except for the one with the linear piece along $x$. Denoting this scalar by $\psi$, we write
\begin{equation}
	\delta \psi =  e^{- i \omega t} \alpha^{-1} \chi(r). 
\end{equation}
The linearized equations of motion that yield non-trivial information are Maxwell's equation, the Klein-Gordon equation for 
$\psi$ and the $(r,x)$ component of Einstein's equation.
These can be written, respectively, as
\begin{align}
\label{eq ax gen d}
a_x'' + \left[ \frac{f'}{f} + \frac{(d-3)}{r} \right] a_x' + \frac{\omega^2}{f^2} a_x  + \frac{\mu (d-2)}{f} \frac{r_0^{d-2}}{r^{d-3}} H_{tx}' &= 0, \\
\label{eq chi gen d}
\chi'' + \left[ \frac{f'}{f} + \frac{(d-1)}{r} \right] \chi' + \frac{\omega^2}{f^2} \chi - \frac{i \omega \alpha^2}{f^2} H_{tx} &= 0, \\
\label{eq Htx gen d}
\frac{i \omega r^2}{f} H_{tx}' + \frac{i \omega \mu (d-2) }{f} \frac{r_0^{d-2}}{r^{d-1}} a_x  -  \chi' &= 0.
\end{align}
%
It is convenient to eliminate $H_{tx}$ by taking a radial derivative of \eqref{eq chi gen d} and plugging the resulting equation for  $H_{tx}'$ 
in \eqref{eq ax gen d} and \eqref{eq chi gen d}. Then, introducing 
\begin{equation}\label{def phi gen d}
	\phi = \omega^{-1} r^{d-1} f  \chi', 
\end{equation}
\noindent we find
\begin{align}
\label{eq2 ax gen d}
 r^{3-d}(r^{d-3} f a_x')' +  \frac{\omega^2}{f} a_x &=  (d-2)^2 \mu^2 \frac{r_0^{2(d-2)}}{r^{2(d-1)}} a_x + i (d-2) \mu \frac{r_0^{d-2}}{r^{2(d-1)}} \phi, \\
\label{eq phi gen d}
	r^{d-1}(r^{1-d} f \phi')' + \frac{\omega^2}{f} \phi  &= - i (d-2) \alpha^2 \mu \frac{r_0^{d-2}}{r^2} a_x + \frac{ \alpha^2}{r^2}  \phi. 
\end{align}
Closely following \cite{Blake:2013bqa}, we now proceed to derive the DC conductivity of our model. 
The key observation is that, as noted previously in \cite{Iqbal:2008by}, there exists a massless mode which can easily be established noting that the mass matrix associated to \eqref{eq2 ax gen d}, \eqref{eq phi gen d} has a vanishing determinant. 
Moreover, taking the linear combinations
\begin{align}
	\lambda_1 &= B(r)^{-1} \left(  a_x - \frac{i (d-2) \mu r_0^{d-2}  }{  \alpha^2 r^{2(d-2)}  }   \phi \right), \\
	\lambda_2 &= B(r)^{-1} \left( \frac{(d-2)^2 \mu^2 r_0^{2(d-2)}}{\alpha^2 r^{2(d-2)}}   a_x + i \frac{(d-2) \mu r_0^{d-2}}{ \alpha^2 r^{2(d-2)} } \phi \right),
\end{align}
\noindent where
\begin{equation}
	B(r) = \left( 1 + \frac{ (d-2)^2 \mu^2 r_0^{2(d-2)} }{ \alpha^2 r^{2(d-2)}   } \right),
\end{equation}
\noindent makes it explicit that $\lambda_1$ is massless. In fact, its equation of motion reads
\begin{equation}\label{eq lambda1}
	\left(  r^{d-3} f B \lambda_1' - 2 (d-2) r^{d-4} f \lambda_2 \right)' + \frac{\omega^2 B}{f r^{3-d}} \lambda_1  = 0.
\end{equation}
It follows that for $\omega = 0$ the following quantity is radially conserved 
\begin{equation}\label{def Pi}
	\Pi = r^{d-3} f B \lambda_1' - 2 (d-2) r^{d-4} f \lambda_2.
\end{equation}

In order to compute the conductivity we turn on a source for the current and set the source for the dual operator to $\chi$ to zero. Taking into account the definition \eqref{def phi gen d}, we conclude that  to order $O(\omega^2)$ the asymptotics of the fields are given by,
\begin{align}
	\phi &=   \frac{\alpha}{\omega} \langle O \rangle e^{i\omega t}  + \ldots, \\
	a_x &= a_x^{(0)} + \frac{ \langle J_x \rangle e^{i\omega t}
}{(d-2) r^{d-2}} + \ldots
\end{align}
Note that the log terms that appear in odd bulk dimensions make an $O(\omega^2)$ contribution. The conductivity is given by the boundary data appearing in these expansions, 
\begin{equation}\label{sigma def}
	\sigma(\omega) = \frac{ \langle J_x \rangle }{i \omega \delta A_x^{(0)}  },
	\end{equation}
where $\delta A_x^{(0)} = e^{-i \omega t} a_x^{(0)}$.
The DC conductivity is defined as the limit of \eqref{sigma def} as $\omega$ goes to zero, 
\begin{equation}
	\sigma_{DC} = \lim_{\omega \to 0} \sigma(\omega).
\end{equation}
In odd bulk dimensions, there is an ambiguity associated to the choice of renormalization scheme which reflected in the appearance of logarithms in the boundary expansion for $a_x$\footnote{Note that massless scalar and graviton fields also contain logarithms in the near boundary expansion for odd bulk dimensions.}. 
However, as mentioned above, these log terms vanish to order $O(\omega^2)$ so they can be ignored in the computation of the DC conductivity.

As an auxiliary quantity, we define
\begin{equation}\label{sigma DC of r}
	\sigma_{DC}(r) = \lim_{\omega \to 0} \frac{-\Pi}{i \omega \lambda_1} \bigg |_{r}.
\end{equation}
Note that $\sigma_{DC}(\infty) = \sigma_{DC}$. The crucial point of the derivation is that \eqref{sigma DC of r} is actually $r$ independent, so that it can be evaluated at the horizon. Our equations of motion have exactly the same structure as those in \cite{Blake:2013bqa}, so the proof of $\sigma'(r)_{DC} = 0$ is almost identical to the one found in that reference. For completeness, we review it in appendix \ref{sigma DC const}. 
To evaluate \eqref{sigma DC of r} at the horizon $r = r_0$, we use that the fields satisfy in going boundary conditions there, namely
\begin{align}
\nonumber
	\phi &= (r-r_0)^{- i \omega/f'(r_0)}[ \phi^H + O( (r-r_0))  ], \\
\label{near H gen d}
	a_x &= (r-r_0)^{- i \omega/f'(r_0)}[ a_x^H + O( (r-r_0))  ].
\end{align}
Plugging \eqref{near H gen d} in \eqref{sigma DC of r} and setting $r = r_0$ we find
\begin{equation}\label{sigma DC gen d final}
	\sigma_{DC} = r_0^{d-3}\left( 1 + (d-2)^2 \frac{\mu^2}{\alpha^2}\right).
\end{equation}
As promised, the DC conductivity has a finite value in agreement with the physics of momentum dissipation. Note that the result \eqref{sigma DC gen d final} is the DC conductivity for any of the black holes in the family parameterised by $T, \mu$ and $\alpha$. 

Moreover, for $d = 3$, expression \eqref{sigma DC gen d final} coincides with the result in massive gravity provided we identify $\alpha$ with the graviton mass commonly denoted $m_\beta$ (with the other mass parameter of massive gravity usually considered in this context, $m_\alpha$, set to zero).\footnote{Specifically for \cite{Blake:2013bqa} we have $\beta_{there} = -\alpha_{here}/2$ and $\alpha_{there}=0$ as can be seen from the background metric.}  As mentioned in the introduction, the correspondence between the model presented in this paper and that particular sector of massive gravity goes largely beyond this point, as we shall establish in section \ref{sec: comp MG}.

Associated to the finite DC conductivity is a scattering timescale, $\tau$. We have not computed this for general dimensions, but for the case of $d=3$ we recover the massive gravity value \cite{Davison:2013jba}. The relation of this theory in $d=3$ to massive gravity is discussed in more detail in section \ref{sec D=4}.

In the case $d=3$ at fixed $\alpha/\mu$ the DC conductivity is temperature independent. The key difference as we move to $d\neq 3$ is the temperature dependence, following from factors of $r_0$ which appear in order to make up the dimension. At low temperatures the behaviour is, 
\be
\frac{\sigma_{DC}}{\mu^{d-3}} = a_d\left(\tfrac{\alpha}{\mu}\right) + b_d\left(\tfrac{\alpha}{\mu}\right) \frac{T}{\mu} + O\left(\frac{T}{\mu}\right)^2, \qquad  T\ll \mu,
\ee
where $a_d$ and $b_d$ are coefficients which depend on the dimension and $\alpha/\mu$, in particular $b_3=0$. At large temperatures we have,
\be
\frac{\sigma_{DC}}{\mu^{d-3}} = c_d\left(\tfrac{\alpha}{\mu}\right)\left(\frac{T}{\mu}\right)^{d-3}\left(1+ O\left(\frac{T}{\mu}\right)^{-2}\right), \qquad T\gg\mu,
\ee
which follows on dimensional grounds. 

Before closing this section, we mention that the system \eqref{eq2 ax gen d}, \eqref{eq phi gen d} can be in fact decoupled by introducing the master fields $\Phi_\pm$ which are defined as
\begin{equation}
	\phi = r^{d-2} (c_+ \Phi_+ + c_- \Phi_- ) \qquad a_x = - i ( \Phi_+ + \Phi_-  ),
\end{equation}
\noindent where
\begin{equation}
	c_\pm = \frac{1}{2 \mu r_0^{d-2}} \left[  d m_0 \pm \left( d^2 m_0^2 + 4 r_0^{2(d-2)} \mu^2 \alpha^2  \right)^{1/2}   \right].
\end{equation}

The master fields are governed by the equations
\begin{equation}
	r^{5-d} ( f r^{d-3} \Phi_\pm' )' + \left(  \frac{r^2 \omega^2}{f}  - \frac{(d-2)^2 \mu^2 r_0^{2(d-2)} }{ r^{2(d-2)}  }   
	+c_\pm  (d-2) \mu \frac{ r_0^{d-2} }{ r^{d-2}  } \right) \Phi_\pm = 0.
\end{equation}

\section{The case of $d=3$\label{sec D=4}}
In this section we study more closely the four dimensional version of the proposed model of momentum dissipation. The purpose of doing so is twofold. Firstly, fixing the number of dimensions facilitates the holographic renormalization procedure \cite{Henningson:1998gx,deHaro:2000xn}, which will allow us to explicitly obtain the Ward identity \eqref{ward cons general} and check that, at the linear level in the fluctuations, momentum in the dual field theory is not conserved. Secondly, specializing to $d= 3$ will permit us to draw a comparison with massive gravity,  finding striking practical similarity despite the conceptual differences.

\subsection{Holographic renormalisation}
\label{sec: HR d=3}
A standard analysis reveals the required counterterms and lead us to the renormalised action, 
\begin{equation}\label{S ren d=3}
	S_{ren} = S_0 + \bi \sqrt{-\gamma} \left( -4 - R[\gamma] - \frac{1}{2}  \psi_I \Box_\gamma \psi_I \right).
\end{equation}
where $\gamma$ is the induced metric on the boundary, see e.g. \cite{Bianchi:2001kw} and appendix \ref{app:HR} for details.
 From this we may derive the stress tensor via, 
\begin{align}\label{renT}
	\left<T^{ij}\right> &= \lim_{r\to\infty} \; \frac{2}{r^{-5} \sqrt{-\gamma}}\frac{\delta S_{ren}}{\delta\gamma_{ij}} \\
								&= 	\lim_{r\to\infty} r^5 \left[ 2 \left(K^{ij} - K \gamma^{ij} + G^{ij}_\gamma -2 \gamma^{ij}\right) + \frac{1}{2}\gamma^{ij} \partial\psi_I \cdot \partial\psi_I - \nabla^i\psi_I \nabla^j \psi_I\label{renT} \right].
\end{align}
Similarly for the current we have, 
\be
\left<J^i\right> =\lim_{r\to\infty}  \frac{1}{r^{-3}\sqrt{-\gamma}} \frac{\delta S_{ren}}{\delta A_i} =\lim_{r\to\infty} -r^3\; n^\mu F_{\mu}^{~i}. \label{renJ}
\ee
Finally we consider taking variations of the scalars. Recall that the scalar fields will be given spatially dependent sources. Variations of the action with respect to the value of the scalar fields at the boundary give the local scalar vev, in general dependent on boundary coordinates,
\begin{equation}\label{renO}
	\left<O_I\right>= \lim_{r\to\infty}\frac{1}{r^{-3}\sqrt{-\gamma}}\frac{\delta S_{ren}}{\delta \psi_I} =  - \lim_{r\to\infty}r^3\;(n^\mu \partial_\mu \psi_I +  \Box_\gamma \psi_I). 
\end{equation}

It follows from diffeomorphism invariance of $S_{ren}$ that, 
\be
\nabla_j \langle T^{ij} \rangle = \langle O_I \rangle \nabla_i \psi^{(0)}_I +  F^{(0)}_{ij} \langle J^j \rangle 
\ee
and is a straightforward exercise to verify using the bulk equations of motion given the expressions above, see appendix \ref{app:HR}.

We also note at this point that the boundary variations \eqref{renT},\eqref{renJ} and \eqref{renO} do not constitute the complete set of variations for this action. Due to the linear spatial dependence of the scalar sources used in this paper, we also find an additional contribution from boundary terms evaluated at spatial infinity. Whilst they do not contribute to the one-point functions, they will be discussed in the next section where we will see that they contribute to the first law. 

Finally, we specialise the results of this section to the black hole solutions of section \ref{blackbranes}, where we find the non-vanishing components of the stress tensor and current, 
\be
\left<T^{tt}\right> = 2m_0, \quad \left<T^{xx}\right> =\left<T^{yy}\right> =m_0, \qquad \left<J^t\right> = \mu r_0,
\ee 
with $\left<O\right> = 0$.

\subsection{Thermodynamics}

For this section we continue to Euclidean time through, 
\be
t = -i\tau, \qquad I_{ren} = -i S_{ren}.
\ee
where $S_{ren}$ is the renormalised on-shell action \eqref{S ren d=3}. Evaluating the Euclidean on-shell action in the case of the $d=3$ solutions of section \ref{blackbranes} we find, 
\be
I_{ren} = \beta w V_2,
\ee
where $V_2 = \int d^2x$ is the infinite spatial volume of the boundary, $\beta = \Delta \tau$ is the inverse temperature and $w$ is the thermodynamic potential per unit volume, which satisfies
\be
w = \varepsilon -Ts - \mu \rho\label{free1}.
\ee
where $\varepsilon = \langle T^{tt} \rangle$, $\rho = \langle J^t \rangle$ and $T$, $s$ are given in \eqref{T gen d}, \eqref{s gen d} with $d = 3$.
We must be a little careful when deriving the pressure because the scalar fields depend on the boundary coordinates. In particular, 
\be
p \neq \left<T^x_x\right>.\label{pressureclaim}
\ee
This is similar to the situation where one applies a constant magnetic field \cite{Hartnoll:2008kx, Donos:2012yu}, and we proceed in analogy with this case. To understand \eqref{pressureclaim}, it is instructive to consider a finite portion of our system of volume $V$. We may calculate the pressure by varying the total energy $E=\varepsilon V$ at fixed $S = s V$, $Q = \rho V$, and in this case at fixed $\alpha$ too,
\be
p=-\frac{\partial E}{\partial V}\bigg|_{S,Q,\alpha} = \left<T^x_x\right>+ r_0 \alpha^2 .
\ee
and moreover, $w =-p$.\footnote{If we had altered the value of $\alpha$ during the volume variation such that $\alpha^2V$ remained fixed (thus keeping the scalar field fixed on the boundary of $V$) we instead find the component of the stress tensor $\left<T^x_x\right>$.}  

Next we may consider variations of the Euclidean on-shell action with respect to $T, \mu$ and $\alpha^2$.  We find, 
\be
\delta w = -s \delta T - \rho \delta \mu - k_\alpha \delta (\alpha^2)\label{firstlaw0},
\ee
where we have allowed for variations in the gradient of the source $\alpha$, introducing its conjugate $k_\alpha$.  This quantity is analogous to the magnetisation per unit volume in the case where one applies a magnetic field. Here, $k_\alpha$ is given by,
\be
k_\alpha = r_0.
\ee
Continuing the analogy, we may also define a susceptibility for turning on $\alpha^2$,
\be
\chi_\alpha = \frac{\partial k_\alpha}{\partial \alpha^2}\bigg|_{T,\mu} = \frac{1}{2} \left[ (4 \pi T)^2 + 3 (\mu^2 + 2 \alpha^2)  \right]^{-1/2}.
\ee
Combining \eqref{firstlaw0} and \eqref{free1} we obtain the first law, 
\be
\delta\varepsilon = T\delta s + \mu \delta\rho - k_\alpha \delta (\alpha^2).
\ee
Finally, we may equate the two expressions for the free energy \eqref{free1} and $w=-p$, arriving at the following Smarr-like formula, 
\be
\varepsilon +p = Ts + \mu \rho\label{smarr}.
\ee


\subsection{Shear modes}

In the present section we consider linearized fluctuations with arbitrary spatial and time dependence, which we write as
\begin{align}
\nonumber
\delta g_{ry} &= e^{- i \omega t + i k x} r^2 H_{ry}(r), &  \delta g_{ty} &=  e^{- i \omega t + i k x} r^2 H_{ty}(r), & \delta g_{xy} &= e^{- i \omega t + i k x} r^2 H_{xy}(r), \\
\label{shear def}
	\delta A_y &= e^{- i \omega t + i k x}  a_y(r), & \delta \psi_1 &=  0, &  \delta \psi_2 &=  e^{- i \omega t + i k x} \alpha^{-1} \chi_2 (r).
\end{align}
\noindent Here we have made use of the isotropy of the background metric to align the momentum in the $x$ direction without loss of generality.
Perturbations of the form \eqref{shear def} are usually referred to as transverse or shear modes, due to the fact that the polarizations lie on the plane orthogonal to the momentum. 
All the other components decouple by rotational symmetry. The form of the ansatz above is left invariant by the action of the vector field
\begin{equation}\label{vec gauge shear}
	\xi = e^{- i \omega t + i k x}  \xi^y(r) \partial_y.
\end{equation}
\noindent which acts on the fourier components of the linearized fields as
\begin{equation}
	\delta_\xi H_{ry} = r^2 \xi^y \hs ' , \qquad \delta H_{ty} = - i \omega \xi^y,  \qquad \delta H_{xy} = i k \xi^y, \qquad \delta a_y = 0, \qquad \delta \chi_2 = \alpha^2 \xi^y.
\end{equation}
We will postpone fixing this ambiguity until later on. The equations of motion are
\begin{align}
\label{eq chi d=3}
	\frac{1}{r^2 f} ( r^2 f \chi_2'  )' + \left( \frac{\omega^2}{f^2} - \frac{k^2}{r^2 f} \right ) \chi_2
	- \alpha^2 \left[ \frac{1}{r^2 f} ( r^2 f H_{ry}  )' + \frac{i \omega}{f^2} H_{ty} + \frac{i k}{r^2 f} H_{xy}  \right] &= 0, \\
\label{eq Hxy d=3}
	\frac{1}{r^2 f} ( r^2 f H_{xy}'  )' + \left( \frac{\omega^2}{f^2} - \frac{\alpha^2}{r^2 f} \right ) H_{xy} 
	- i k \left[ \frac{1}{r^2 f} ( r^2 f H_{ry}  )' + \frac{i \omega}{f^2} H_{ty}   - \frac{1}{r^2 f} \chi_2  \right] &= 0, 	\\
\label{eq ay d=3}
	a_y'' + \frac{f'}{f} a_y' + \left( \frac{\omega^2}{f^2} -  \frac{k^2}{r^2 f} \right) a_y + \frac{\mu r_0}{f} ( H_{ty}' + i \omega H_{ry} ) &= 0, \\
\label{eq Hty d=3}
	\frac{i \mu r_0 \omega}{r^2 f} a_y + \frac{i \omega r^2}{f} H_{ty}' + i k H_{xy}' + \left[ (k^2 + \alpha^2) - \frac{r^2 \omega^2}{f}  \right] H_{ry} - \chi_2' &= 0, 
\end{align}
\noindent in addition to the linearly dependent equation
\begin{equation}\label{extra eq Hty d=3}
	r^2 H_{ty}'' + 4 r H_{ty}' - \frac{k^2 + \alpha^2}{f} H_{ty} + i \omega \left(  4 r H_{ry} + \frac{i k}{f} H_{xy} - \frac{1}{f} \chi_2 + r^2 H'_{ry}  \right) + \frac{\mu r_0}{r^2} a_y' = 0.
\end{equation}

Note that, because of the gauge freedom associated to \eqref{vec gauge shear}, we obtain only four independent equations of motion. 
Remarkably, it is possible to fully decouple the fluctuation equations \eqref{eq chi d=3}-\eqref{eq Hty d=3}.  To do so, we introduce the gauge invariant combinations $\Phi_0$, $\Phi_1$, $\Phi_\pm$ via 
\begin{align}
	H_{ry} &= \frac{1}{\alpha^2} \chi_2' -  \frac{\omega}{r f} ( c_+ \Phi_+ + c_- \Phi_- ) - \frac{k}{k^2 + \alpha^2} \Phi_1', \\
\label{def H}
	H_{ty} &= - i \left (\frac{\omega}{\alpha^2} \chi_2 - \Phi_0 \right), \\
	H_{xy} & = i \left ( \frac{k}{\alpha^2} \chi_2 - \Phi_1 \right), \\
	a_y & = - i (k^2 + \alpha^2) ( \Phi_+ + \Phi_-   ),
\end{align}
\noindent where
\begin{equation}\label{c MF}
	c_\pm = \frac{1}{2 \mu r_0} \left[ 3 m_0 \pm (9 m_0^2 + 4 \mu^2 r_0^2 ( k^2 + \alpha^2))^{1/2} \right].
\end{equation}
Then, we can readily show that the equations of motion \eqref{eq chi d=3}-\eqref{eq Hty d=3} reduce to
\begin{align}
\label{MF eq}
	r^2 ( f \Phi_\pm' )' + \left( \frac{r^2 \omega^2}{f} - k^2 - \frac{\mu^2 r_0^2}{r^2} + \frac{\mu r_0}{r} c_\pm \right) \Phi_\pm &= 0, \\
\label{eq Phi1}
		\frac{1}{r^2 f} (r^2 f \Phi_1')' + \left( \frac{\omega^2}{f^2}  -  \frac{(k^2 + \alpha^2)}{r^2 f} \right) \Phi_1 &= 0, \\
\label{eq Phi0}
		\Phi_0 + \frac{f}{r^2} ( c_+ r \Phi_+ + c_- r \Phi_- )'  - \frac{k \omega}{k^2 + \alpha^2} \Phi_1 &=0. 
\end{align}
Note that $\Phi_\pm$ encode the gauge invariant information about $a_y$, and as such, they govern the physics of the conductivity. 
In the next section we will derive the equations for the shear modes in massive gravity, and conclude that the associated master fields $\Phi_\pm$ satisfy an equation identical to \eqref{MF eq}, showing that the conductivities in both theories coincide, even at non-zero momentum.

Let us discuss the computation of the optical conductivity in this model. This is done by setting $k=0$ in \eqref{eq chi d=3}-\eqref{eq Hty d=3}, as a result of  which \eqref{eq Hxy d=3} decouples. Moreover, in order to facilitate the holographic interpretation, we work in radial gauge $H_{ry} = 0$.
Then, the system \eqref{eq chi d=3}-\eqref{eq Hty d=3} reduces to \eqref{eq ax gen d}-\eqref{eq Htx gen d} for $d= 3$. 

For the shear modes, the Ward identity \eqref{ward cons general} reduces to
\begin{equation}\label{linear ward}
	\partial_t \delta \langle P_x \rangle = - \alpha \delta \langle O_1 \rangle  + \rho \partial_t \delta A_x^{(0)},
\end{equation}
\noindent where $\delta \langle P_x \rangle$, $\delta \langle O_1 \rangle$ are the $\delta A_x^{(0)}$ are the linearized contribution to the momentum density $\langle P_x \rangle$, the vev $\langle O_1 \rangle$ and the source $A_x^{(0)} $, respectively. Of course, a similar equation holds for the $y$ component, since at $k=0$ there is no distinction between transverse and longitudinal modes.
We see that at the level of fluctuations non-conservation results when the right hand side of \eqref{linear ward} is finite. In fact, \eqref{linear ward} has the hallmarks of the Drude model. The second term on the right hand side is simply a driving term proportional to the boundary electric field. If in addition we have solutions to the shear mode fluctuation equations with ingoing boundary conditions for which $\delta\langle O_1 \rangle$ is finite, this gives rise to the dissipative piece.  Solutions to these equations with ingoing boundary conditions do exist as we know from the massive gravity results of \cite{Vegh:2013sk,Davison:2013jba}, where it was shown that the stress tensor fluctuations were of Drude form at low frequencies, i.e. with the right hand side of \eqref{linear ward} non-vanishing and proportional to $\delta \langle P_x \rangle$ in the absence of the electric field. This makes it explicit that, while the background boundary geometry is homogeneous and isotropic, momentum does dissipate at the level of fluctuations. The connection with massive gravity is discussed in more detail in section \ref{sec: comp MG}.

\section{Comparison with massive gravity}
\label{sec: comp MG}

The main goal of the present section is to compare in more detail the dynamics of massive gravity and our model by looking at shear fluctuations for finite momentum. These massive gravity modes where studied previously in \cite{Vegh:2013sk} at zero momentum and in \cite{Davison:2013jba} at finite momentum, here we simply recast the results in a way that facilitates the comparison. As a by product of our analysis, we are able to extend the results of \cite{Davison:2013jba} in that we decouple the shear fluctuations for non-zero frequencies and momentum. 

In the reminder of this section we will consider the sector of massive gravity described by the action 
\begin{equation}
	I_{MG} = \int_M \sqrt{g} \left[ R - 2 \Lambda - \frac{1}{4} F^2  + \beta m^2 {\cal L}_\beta   \right ] d^4 x. 
\end{equation}
\noindent where
\begin{equation}\label{L beta}
	{\cal L}_\beta =  [{\cal K}]^2 -  [{\cal K}^2].
\end{equation}
\noindent Here the matrix ${\cal K}^\mu \hs _\nu$ is defined by ${\cal K}^\mu \hs _\alpha {\cal K}^\alpha \hs _\nu = g^{\mu \alpha} f_{\alpha \nu}$, with $f_{\alpha \beta}$ a fixed tensor, commonly referred to as the ``reference metric". The square brackets in \eqref{L beta} denote a matrix trace, e.g. $[{\cal K}] = {\cal K}^\mu \hs _\mu$.
Choosing $f_{\mu \nu} = {\rm diag}(0,0,F,F)$ in the coordinate system $(r,t,x,y)$, we find
\begin{equation}
	{\cal L}_\beta = 2 F [  g^{xx} g^{yy}  - (g^{xy})^2 ]^{1/2}.
\end{equation}
Therefore, Einstein's equations read
\begin{align}
	R_{\mu \nu} - & \frac{1}{2} g_{\mu \nu} R + \Lambda g_{\mu \nu} =  \frac{1}{2}\left( F^\alpha \hs _\mu F_{\alpha \nu}  - \frac{1}{4} g_{\mu \nu} F^2 \right) \\
\nonumber
	& + \frac{m_\beta^2}{[ g^{xx} g^{yy} -  (g^{xy})^2  ]^{1/2}}  (  g^{yy} \delta^x_\mu \delta^x_\nu  + g^{xx}  \delta^y_\mu \delta^y_\nu  - 2 g^{xy} \delta^x_{(\mu} \delta^y_{\nu)}   -  g_{\mu \nu} [ g^{xx} g^{yy} -  (g^{xy})^2  ] ),
\end{align} 
\noindent where $m_\beta^2 = - \beta F m^2$. These are to be supplemented by Maxwell's equations
\begin{equation}
	\partial_\mu (\sqrt{g} F^{\mu \nu}) = 0.
\end{equation}

The system admits the solution
\begin{equation}\label{backgnd anstaz MG}
	ds^2 = - f_{MG}(r) dt^2 + \frac{dr^2}{f_{MG}(r)} + r^2(dx^2 + dy^2), \qquad A = \mu \left(1 - \frac{r_0}{r}\right) dt, 
\end{equation}
\begin{equation}\label{g orig MG}
		f_{MG}(r) = r^2  - m_\beta^2 - \frac{m_0}{r}  + \frac{\mu^2 r^2_0}{4 r^2},
\end{equation}
\noindent which coincides with the $d=3$ case of the background \eqref{gen D ansatz}, \eqref{f gen d}, \eqref{At soln gen d} provided we set $\alpha^2 = 2 m_\beta^2$. Keeping in mind that this identification between the parameters is to be made, we will drop the subscript $MG$ in $f$. \\

We consider shear perturbations which we write as 
\begin{align}
\nonumber
	\delta g_{ry} &= e^{- i \omega t + i k x} r^2 H_{ry}(r),  & \delta g_{ty} &= e^{- i \omega t + i k x} r^2 H_{ty}(r), \\
	\delta g_{xy} &= e^{- i \omega t + i k x} r^2 H_{xy}(r), & \delta A_y    &= e^{- i \omega t + i k x}  a_y(r).
\end{align}
These are governed by the fluctuation equations 
\begin{align}
\label{eq Hxy MG}
	\frac{1}{r^2 f} ( r^2 f H_{xy}'  )' + \frac{\omega^2}{f^2}  H_{xy} 
	- i k \left[ \frac{1}{r^2 f} ( r^2 f H_{ry}  )' + \frac{i \omega}{f^2} H_{ty}  \right ] &= 0, 	\\
\label{eq ay MG}
	a_y'' + \frac{f'}{f} a_y' + \left( \frac{\omega^2}{f^2} -  \frac{k^2}{r^2 f} \right) a_y + \frac{\mu r_0}{f} ( H_{ty}' + i \omega H_{ry} ) &= 0, \\
\label{eq Hty MG}
	\frac{i \mu r_0 \omega}{r^2 f} a_y + \frac{i \omega r^2}{f} H_{ty}' + i k H_{xy}' + \left[ (k^2 + 2 m_\beta^2) - \frac{r^2 \omega^2}{f}  \right] H_{ry} &= 0, \\
\label{eq Hty MG}
	r^2 H_{ty}'' + 4 r H_{ty}' - \frac{k^2 + 2 m_\beta^2}{f} H_{ty} + i \omega \left(  4 r H_{ry} + \frac{i k}{f} H_{xy}  + r^2 H_{ry}  \right) + \frac{\mu r_0}{r^2} a_y' &= 0. 
\end{align}
Fixing the gauge as $\chi_2 = 0$ in the equations for the shear modes \eqref{eq chi d=3}-\eqref{extra eq Hty d=3} makes the comparison with \eqref{eq ay MG}-\eqref{eq Hty MG} 
straightforward. After identifying $\alpha^2 = 2 m_\beta^2$, which follows from the comparison of the background solutions, we note that the only difference between the two system of equations is the term $r^{-2} f^{-1} \alpha^2 H_{xy}$ in \eqref{eq Hxy MG} and \eqref{eq Hxy d=3}. 
In order to draw the parallel on more physical grounds, it is convenient to decouple the fluctuation equations introducing gauge invariant master fields
as in \eqref{def H}, \eqref{c MF} with $\chi_2 = 0$ and $\alpha^2 = 2 m_\beta^2$, obtaining 
\begin{align}
\label{MF eq MG}
	r^2 ( f \Phi_\pm' )' + \left( \frac{r^2 \omega^2}{f} - k^2 - \frac{\mu^2 r_0^2}{r^2} + \frac{\mu r_0}{r} c_\pm \right) \Phi_\pm &= 0, \\
\label{eq Phi1 MG}
		\frac{1}{r^2 f} (r^2 f \Phi_1')' + \frac{\omega^2}{f^2}  \Phi_1 &= 0, \\
\label{eq Phi0 MG}
		\Phi_0 + \frac{f}{r^2} ( c_+ r \Phi_+ + c_- r \Phi_- )'  - \frac{k \omega}{k^2 + 2 m_\beta^2} \Phi_1 &=0.
\end{align}
Note that equation \eqref{MF eq MG} is identical to its counter part \eqref{MF eq}. Because $\Phi_\pm$ contain the gauge invariant information about $a_y$, it follows that the (shear) current-current correlators are exactly those of massive gravity for all frequencies and momenta. 
The form of the algebraic equation for $\Phi_0$ is also the same as
\eqref{eq Phi1}, however, their physical content is not the same since they involve $\Phi_1$ which satisfies different equations in each model, c.f. \eqref{eq Phi1 MG}, \eqref{eq Phi1}. Because $\Phi_0$ carries the gauge invariant information about $H_{tx}$, we see that the thermal conductivity in our model is different to the one obtained in massive gravity. 

Finally we note that in this section we worked directly in fluctuations of metric components. We could equivalently have worked in a St\"uckelberg formulation, introducing a scalar field for each spatial boundary direction. 

\section{Discussion
\label{discussion}}

We presented a simple holographic model for momentum relaxation, exploiting spatially dependent sources for scalar operators. In general turning on such sources would render the bulk solutions inhomogeneous and anisotropic, but by adding $d-1$ massless scalars with sources linear in coordinates we analytically constructed homogeneous and isotropic examples. 

We extracted the finite DC optical conductivity, $\sigma_{DC}$, analytically for general values of the parameters in general dimensions. In the case of $d=3$ at fixed gradient $\alpha$ and chemical potential $\mu$, $\sigma_{DC}$ is temperature independent. Moving away from $d=3$ reveals temperature dependence in accordance with the dimension of the optical conductivity.

In the case of four bulk dimensions, the homogeneous and isotropic black brane solutions are the same as those constructed by \cite{Vegh:2013sk} for the nonlinear massive gravity theory proposed in \cite{deRham:2010kj}. 
Moreover, we have analysed the shear mode fluctuations finding that whilst the modes controlling the thermal conductivities are different in each theory, the equations governing the current-current correlators are identical, and consequentially the electric conductivity is the same.\footnote{At the qualitative level, one may be able to understand this similarity by equivalently thinking of massive gravity in the St\"uckelberg field formulation. In this case we would have Einstein gravity coupled to massless scalar fields with non-standard kinetic terms.}

The field theory interpretation of the physics associated to massive gravity in the bulk is not clear. This is partly due to the fact that, for generic values of the parameters, massive gravity may contain ghosts (see \cite{deRham:2010kj} for the discussion of healthy scenarios).  
The present calculations illustrate that at least the conductivity of the background may be understood as simply turning on particular spatially dependent sources for scalar operators in a bulk theory with standard matter content.  
Since the masses of the fields are above the Breitenlohner-Freedman bound \cite{Breitenlohner:1982jf} and the boundary conditions are such that the dual operators respect the relevant unitarity bounds, we do not expect violations of unitarity in our model, which then provides a simple well-defined setup to discuss the phenomenology of momentum relaxation in holography.  
Moreover, since the matter fields used are relatively standard, it may be possible to find string theory embeddings which would further elucidate the field theory interpretation.  

It is interesting to note that because our scalar sources are linear in coordinates, the system does not seem to describe a lattice. This suggests in the case of $d=3$ that the physics seen in the nonlinear massive gravity case of \cite{deRham:2010kj,Vegh:2013sk} is not lattice physics.\footnote{We thank Marika Taylor for raising this point.} An intermediate frequency scaling regime for numerical lattice set-ups was observed in \cite{Horowitz:2012ky,Horowitz:2012gs}. Indeed, such a robust regime was not found in this model, nor was it found in the massive gravity case \cite{Vegh:2013sk}.

Unlike in the case of massive gravity, it is clear how we can consistently modify or extend the proposed model. Indeed, in this paper we considered the case of arbitrary dimensions. It would be interesting to investigate different backgrounds such as those with vanishing entropy at low temperature along the lines of \cite{Davison:2013txa}. Finally we note that a natural extension of the current model presents itself -- there are additional bulk terms we can add at two-derivative order which respect the shift-symmetry of the scalars, $\int\psi_I F\wedge F$. This leads to a bulk Einstein-Maxwell model with $d-1$ axion fields. 

\section*{Acknowledgements}

We thank Richard Davison, Veronika Hubeny, Don Marolf, Mukund Rangamani, Simon Ross, Kostas Skenderis, Gianmassimo Tasinato and Marika Taylor for useful discussions. We would additionally like to thank Marco Caldarelli, Jerome Gauntlett and especially Simon Gentle for useful comments and discussions on v1 of this paper.
TA is supported by STFC by the National Science Foundation under Grant No. NSF PHY11-25915 and Grant No PHY08-55415.

\appendix

\section{$\sigma_{DC}(r)$ is constant}
\label{sigma DC const}

In this appendix we show that the a priori radially dependent conductivity is in fact independent of the slice, closely following \cite{Blake:2013bqa}.
Computing the radial derivative of \eqref{sigma DC of r}, we obtain
\begin{equation}
	\sigma_{DC}'(r) = \lim_{\omega \to 0} \frac{1}{i \omega} \left[  \frac{\Pi'}{\lambda_1} - \left( \frac{\Pi}{\lambda_1} \right) \left( \frac{\lambda_1'}{ \lambda_1 } \right)   \right].
\end{equation}
\noindent The desired result follows if we are able to show that $\Pi'/\lambda_1 \sim O(\omega^2)$, $\Pi/ \lambda_1 \sim \lambda_1' / \lambda_1 \sim O(\omega)$ for small $\omega$. The first statement is a trivial consequence of equation \eqref{eq lambda1}. Moreover, we have that $\Pi/ \lambda_1 \sim O(\omega)$ since this property is true at the horizon due to ingoing boundary conditions and it is preserved into the bulk by \eqref{eq lambda1}\footnote{We can imagine discretizing \eqref{eq lambda1} obtaining $\Pi(r_0 + \epsilon) \sim \Pi(r_0) + \epsilon O(\omega^2)$ which indeed yields $\Pi(r_0 + \epsilon) \sim \Pi(r_0) \sim O(\omega)$.}. Finally, we can schematically write the equation for $\lambda_2$ as
\begin{equation}\label{schem eq lambda2}
	L_2 \lambda_2 + p(r) \lambda_2 + \omega^2 q(r) \lambda_2 \sim \Pi, 
\end{equation}
\noindent where $L_2$ is a linear differential operator and $p$ and $q$ are functions of $r$ with no $\omega$ dependence. The only feature of \eqref{schem eq lambda2} that is relevant for us is that, due to the masslessness of $\lambda_1$, it only couples to $\lambda_2$ via $\Pi$. Now, equation \eqref{schem eq lambda2} yields $\lambda_2 \sim \Pi$ which combined with \eqref{def Pi} implies $\lambda_1' / \lambda_1 \sim O(\omega)$, as desired. 

\section{Details of holographic renormalisation}
\label{app:HR}

In this appendix we record some useful results to carry out the holographic renormalisation procedure. We begin with the metric in Fefferman-Graham from 
\begin{equation}\label{ds2 FG}
	ds^2 =  \frac{d \rho^2}{\rho^2} + \frac{1}{\rho^2} g_{ij}(\rho) dx^i dx^j, 
\end{equation}
\noindent where the conformal boundary is at $\rho = 0$ and choose radial gauge for the connection $A_\rho = 0$. Recall that $i,j$ label all boundary indices. 

\subsection{Asymptotic solution}

The equations of motion read
\begin{align}
\label{Eins rho rho}
	- \frac{1}{2} Tr (g^{-1} g'') + \frac{1}{2 \rho} Tr (g^{-1} g') &+ \frac{1}{4} Tr (g^{-1} g' g^{-1} g') =  
	\frac{\psi_I' \psi_I' }{2}    +\frac{\rho^2}{4} \left[  g^{ij} A'_i A'_j - \frac{1}{2} g^{ik} g^{jl} F_{ij} F_{kl}  \right],	
\end{align}
\begin{equation}\label{Eins i rho}
	\frac{1}{2} g^{jk} ( \nabla_j g'_{ki} - \nabla_i g_{jk}' )  = \frac{1}{2}  \psi'_I \partial_i \psi_I  + \frac{1}{2} \rho^2 F_{ik} g^{kj} A_j',
\end{equation}
\begin{align}
\nonumber
	& R_{ij}   - \frac{1}{2} g_{ij}'' + \frac{1}{2 \rho} [ 2 g_{ij}' + Tr(g^{-1} g') g_{ij} ] - \frac{1}{4} Tr(g^{-1} g') g_{ij}' + \frac{1}{2} (g' g^{-1} g')_{ij}  = \\
\label{Eins ij}
&\frac{1}{2}  \partial_i \psi_I \partial_j \psi_I  + 
	\frac{\rho^2}{2} \left[ A_i' A_j' + g^{kl} F_{ik} F_{jl}  - \frac{g_{ij}}{4} (  2 g^{kl} A_k' A_l' + g^{mn} g^{kl} F_{mk} F_{nl}   ) \right],
\end{align}
\begin{align}
\label{Maxw rho}
	\partial_i( \sqrt{g} g^{ij} A_j' ) &= 0, \\
\label{Maxw i}
	\partial_\rho (  \sqrt{g} g^{ij} A_j'  ) + \partial_k (  \sqrt{g} g^{kl} g^{ij} F_{lj} ) &= 0, \\
\label{KG decomp}
	\partial_\rho (\rho^{-2} \sqrt{g} \partial_\rho \psi_I) + \rho^{-2} \partial_i (  \sqrt{g} g^{ij} \partial_j \psi_I )  &= 0,
\end{align}
\noindent where all indices are raised and lowered with $g_{ij}(\rho)$, which is also used to take the traces. Studying the equations of motion near the boundary we find 
\begin{align}
	g_{ij} &= g^{(0)}_{ij} + \rho^2 g^{(2)}_{ij} + \rho^3 g^{(3)}_{ij} + \ldots, \\
	A_i   &= A^{(0)}_i + \rho A^{(1)}_i  + \ldots,\\
  \psi_I &= \psi^{(0)}_I  + \rho^2 \psi^{(2)}_I + \rho^3 \psi^{(3)}_I + \ldots
\end{align}

Here leading terms $g^{(0)}_{ij}$, $A^{(0)}_i$, $\psi^{(0)}_I$ are arbitrary and interpreted as sources for dual field theory operators corresponding to the stress tensor $T_{ij}$, a global $U(1)$ current $J^i$ and a scalar operator $O_I$, which are in turn proportional to $g^{(3)}_{ij}$, $A^{(1)}_i$, $\psi^{(3)}_I$, respectively. 
The vevs are left largely unrestricted by the asymptotic equations, with the important exception of the relations
\begin{equation}\label{ward seed d}
	\nabla_{(0)}^i A_i^{(1)} = 0,	\qquad Tr (g^{(3)}) = 0, \qquad \nabla_{(0)}^j g^{(3)}_{ij} =  \psi^{(3)}_I \partial_i \psi^{(0)}_I  + \frac{1}{3} F^{(0)}_{ij} A^{(1) j}.
\end{equation}
Here $\nabla_{(0)}^i$ denotes the covariant derivative compatible with $g^{(0)}_{ij}$.
In addition, we find 
\begin{align}
\label{psi 2}
	 \psi^{(2)}_I &=  \frac{1}{2} \Box_0 \psi^{(0)}_I, \\
\label{g2}
	g^{(2)}_{ij} &= - \left ( R^{(0)}_{ij} - \frac{1}{4} g^{(0)}_{ij} R^{(0)} \right) + \frac{1}{2} \left[  \partial_i \psi^{(0)}_I \partial_j \psi^{(0)}_I  - \frac{1}{4} g^{(0)}_{ij} ( g^{(0) kl} \partial_k \psi^{(0)}_I \partial_l \psi^{(0)}_I )    \right],
\end{align}
\noindent where $\Box_0 = g^{(0)}_{ij}\nabla_{(0)}^i \nabla_{(0)}^j$.

\subsection{Variation}

With the asymptotic solution at hand, and noting that the outward pointing unit normal is $n = - \rho^{-1} d \rho$, we can readily check that the on-shell variation of the renormalized action
\begin{equation}\label{S ren d=3 app}
	S_{ren} = S_0 - \bi \sqrt{-\gamma} \left( 4 + R[\gamma] + \frac{1}{2}  \psi_I \Box_\gamma \psi_I \right).
\end{equation}
\noindent yields
\begin{equation}\label{d I ren app}
	\delta S_{ren} = \bi \sqrt{- g^{(0)}} \left[ \frac{3}{2}  g^{(3) ij} \delta g^{(0)}_{ij} + 3 \psi^{(3)}_I \delta \psi^{(0)}_I +  A^{(1) i} \delta A^{(0)}_i  \right].
\end{equation}
 Therefore, we can write the one point functions as
\begin{equation}\label{vevs app}
	\langle T^{ij} \rangle = 3 g^{(3) ij}, \qquad \langle O_I \rangle = 3 \psi^{(3)}_I,  \qquad \langle J^i \rangle = A^{(1) i}.
\end{equation}
By virtue of \eqref{ward seed d}, we find the Ward identities
\begin{align}
\label{current cons app}
 \nabla_i \langle J^i \rangle &= 0, \\
\label{traceless app}
 \langle T^i \hs _i \rangle &= 0, \\
\label{div T app}
	\nabla^j \langle T_{ij} \rangle &= \langle O_I \rangle \nabla_i \psi^{(0)}_I +  F^{(0)}_{ij} \langle J^j \rangle.
\end{align}
Consider a boundary diffeomorphism generated by a vector field of the form $\xi^i = \xi^i (x^j)$, $\xi^\rho = 0$.  The action of these on the sources is
\begin{equation}\label{lie bndy}
	\delta g^{(0)}_{ij} = \nabla_{(0)i} \xi_j + \nabla_{(0)j} \xi_i, \qquad \delta \psi^{(0)}_I = \xi^i \partial_i \psi^{(0)}_I,
	\qquad \delta A_i^{(0)} = \xi^j \partial_j A^{(0)}_i + A_j^{(0)} \partial_i \xi^j.
\end{equation}
Combining \eqref{d I ren app}, \eqref{vevs app}, \eqref{lie bndy}, along with diffeomorphism invariance of the renormalised action implies the Ward identity \eqref{div T app}, as noted in \cite{Bianchi:2001kw}. Similarly, current conservation \eqref{current cons app} is in one to one correspondence with the $U(1)$ transformation $\delta A^{(0)}_i = \nabla_i \Lambda$ while the tracelessness condition \eqref{traceless app} follows from invariance under constant Weyl transformations of the boundary metric implemented by $\xi^\mu = \delta^\mu_\rho \sigma \rho$ where $\sigma$ is a constant.

\bibliography{relax}{}
\bibliographystyle{JHEP-2}

\end{document}